\documentclass{optica-article}

\journal{oe}

\usepackage{subcaption}
\usepackage{tikz}
\usepackage{graphicx}    
\usepackage{multirow}
\usepackage{booktabs}
\usepackage{appendix}
\usepackage{placeins}
\usepackage{indentfirst}
\usepackage{caption} 
\usepackage{array}
\usepackage{braket}
\usepackage{amsmath}
\usepackage[marginal]{footmisc}
\usepackage{bm,color,bbm}

\usepackage{float}
\newcolumntype{.}{D{.}{.}{-1}}
\usepackage{amsopn}

\articletype{Research Article}
\usepackage{hyperref}	
\hypersetup{
	colorlinks = true,
	citecolor = {blue},
	linkcolor = {red}
}
\usepackage{lineno}

\begin{document}
	
	\title{Field-Trial Quantum Key Distribution with Qubit-Based Frame Synchronization}
	
	\author{Rui Guan,\authormark{1} Jingchun Yu,\authormark{1} Zhaoyun Li,\authormark{2} Hongbo Xie,\authormark{3} Yuxing Wei,\authormark{4} Sen Li,\authormark{4} Jing Wen,\authormark{4} Xiaodong Liang,\authormark{2} Yanwei Li,\authormark{3}  and Kejin Wei\authormark{1}}
	
	\address{\authormark{1} Guangxi Key Laboratory for Relativistic Astrophysics, School of Physical Science and Technology, Guangxi University, Nanning 530004, China\\
		\authormark{2} Guangxi Key Laboratory of Optical Network and Information Security, The 34th Research Institute of China Electronics Technology Group Corporation, Guilin 541004, Guangxi, China.\\
		\authormark{3} Ji Hua Laboratory, Foshan City, Guangdong Province, China.\\
		\authormark{4} Guangxi Zhuang Autonomous Region Information Center, Nanning 530201, China.\\
	}

	\begin{abstract}	
	Quantum key distribution (QKD) is a cryptographic technique that uses quantum mechanical principles to enable secure key exchange, offering information-theoretic security guaranteed by physical laws. Practical deployment of QKD requires robust, cost-effective systems that can operate in challenging field environments. A major challenge is achieving reliable clock synchronization without adding hardware complexity. Conventional approaches often use separate classical light signals, which increase costs and introduce noise that degrades quantum channel performance. To address this limitation, we demonstrate a QKD system incorporating a recently proposed qubit-based distributed frame synchronization method, deployed over a metropolitan fiber network in Nanning, China. Using the polarization-encoded one-decoy-state BB84 protocol and the recently proposed qubit-based distributed frame synchronization method, our system achieves synchronization directly from the quantum signal, eliminating the need for dedicated synchronization hardware. Furthermore, to counteract dynamic polarization disturbances in urban fibers, the system integrates qubit-based polarization feedback control, enabling real-time polarization compensation through an automated polarization controller using data recovered from the qubit-based synchronization signals. During 12 hours of continuous operation, the system maintained a low average quantum bit error rate (QBER) of $1.12 \pm 0.48\%$, achieving a secure key rate of 26.6 kbit/s under 18 dB channel loss. Even under a high channel loss of 40 dB, a finite-key secure rate of 115 bit/s was achieved. This study represents a successful long-term validation of a frame-synchronization-based QKD scheme in a real urban environment, demonstrating exceptional stability and high-loss tolerance, and offering an alternative for building practical, scalable, and cost-efficient quantum-secure communication networks.
	\end{abstract}
	
	\section{Introduction}
	Quantum key distribution (QKD) is a cryptographic technique that leverages the fundamental principles of quantum mechanics, such as the quantum no-cloning theorem and the uncertainty principle, to enable two remote parties to generate and share a secret key with information-theoretic security~\cite{1984BENNETT}. Since the introduction of the pioneering BB84 protocol in 1984, QKD has evolved from a theoretical concept into a practical technology~\cite{Orieux2016,2020Pirandola,2020Xu-review,2022Liu,Cao2022,Granados2025}, with significant potential to secure critical communications against current and future threats, including quantum computers.
	
	Currently, most high-performance QKD systems operate in highly controlled laboratory environments~\cite{2018Boaron,2020grunenfelder,2020Wei,2021paraiso,2023zhou,2023grunenfelder,2023li,2023Liu,2023chen}, which differ substantially from real-world urban fiber networks. Transitioning QKD from controlled laboratory demonstrations to robust, cost-effective systems capable of operating in demanding real-world environments poses critical engineering challenges~\cite{Diamanti2016,2019Bacco,2021Liu,2022Chen,Stanley2022,2023LiWei,2023Kao,2023Tang,2024Bertaina,2024ZhouLai,2024Li,Zhu2024,2025Picciariello}. For wide deployment in existing telecommunications infrastructure, such as metropolitan fiber networks, QKD systems must maintain long-term stability despite environmental fluctuations, including temperature changes and physical fiber disturbances.
	
	A key challenge in practical deployment is achieving high-precision clock synchronization between the communicating parties (Alice and Bob) without adding hardware complexity or cost~\cite{Pljonkin2017,meda2025qkd,Krause2025}. Conventional approaches often use separate classical optical pulses transmitted over the same or parallel fiber~\cite{2019Bacco,Chen2021NATUREPHOTONICS,2023LiWei,2024ZhouLai,Chandra2025}. However, this approach increases the system's component count and cost while introducing extraneous light that can generate noise~\cite{da2014JLT,Aleksic2015OPTICSEXPRESS,Burenkov2023,Ramesh2025OPTICSCOMMUNICATIONS}, potentially reducing the sensitivity of single-photon detectors in the quantum channel and degrading overall system performance.
	
	To address this limitation, qubit-based synchronization has been proposed~\cite{Calderaro2020,Agnesi2020,Avesani2021,cochran2021,Scalcon2022,Azahari2024,2025ChenJLT,2025Huang,Sun2025,lu2025fast}. This method derives synchronization signals directly from quantum states, eliminating the need for separate hardware and its associated drawbacks. A notable advancement is the qubit-based distributed frame synchronization method~\cite{2025ChenJLT}, which provides improved tolerance to channel attenuation, allowing the system to operate continuously without interruption. Its feasibility has been verified in laboratory environments~\cite{2023WeiPhotonRes,huang2024cost,2025chenScienceChinaPhysics,tian2025reference}. However, the long-term stability and practical efficacy of this method had not been validated through extensive testing in dynamic urban fiber networks.
	
	In this study, we demonstrate a QKD system incorporating a recently proposed qubit-based distributed frame synchronization method, deployed over a metropolitan fiber network in Nanning, China. Our system uses the polarization-encoded one-decoy-state BB84 protocol. By integrating qubit-based polarization feedback control, the system actively compensates in real time for dynamic polarization disturbances in field-deployed fibers using data recovered from the qubit-based synchronization signals.
	
	Our experiments show that the system maintained a low average quantum bit error rate (QBER) of \(1.12 \pm 0.48\%\) over 12 hours of continuous operation, achieving a secure key rate of 26.6 kbit/s under 18 dB channel loss. Even under a high channel loss of 40 dB, a finite-key secure rate of 115 bit/s was maintained. This study is a successful long-term validation of a frame-synchronization-based QKD scheme in a real urban environment, demonstrating exceptional stability and high-loss tolerance. Our work provides a promising alternative for building practical, scalable, and cost-efficient quantum-secure communication networks.
	
	\begin{figure*}
		\centering
		\includegraphics[width=0.7\linewidth]{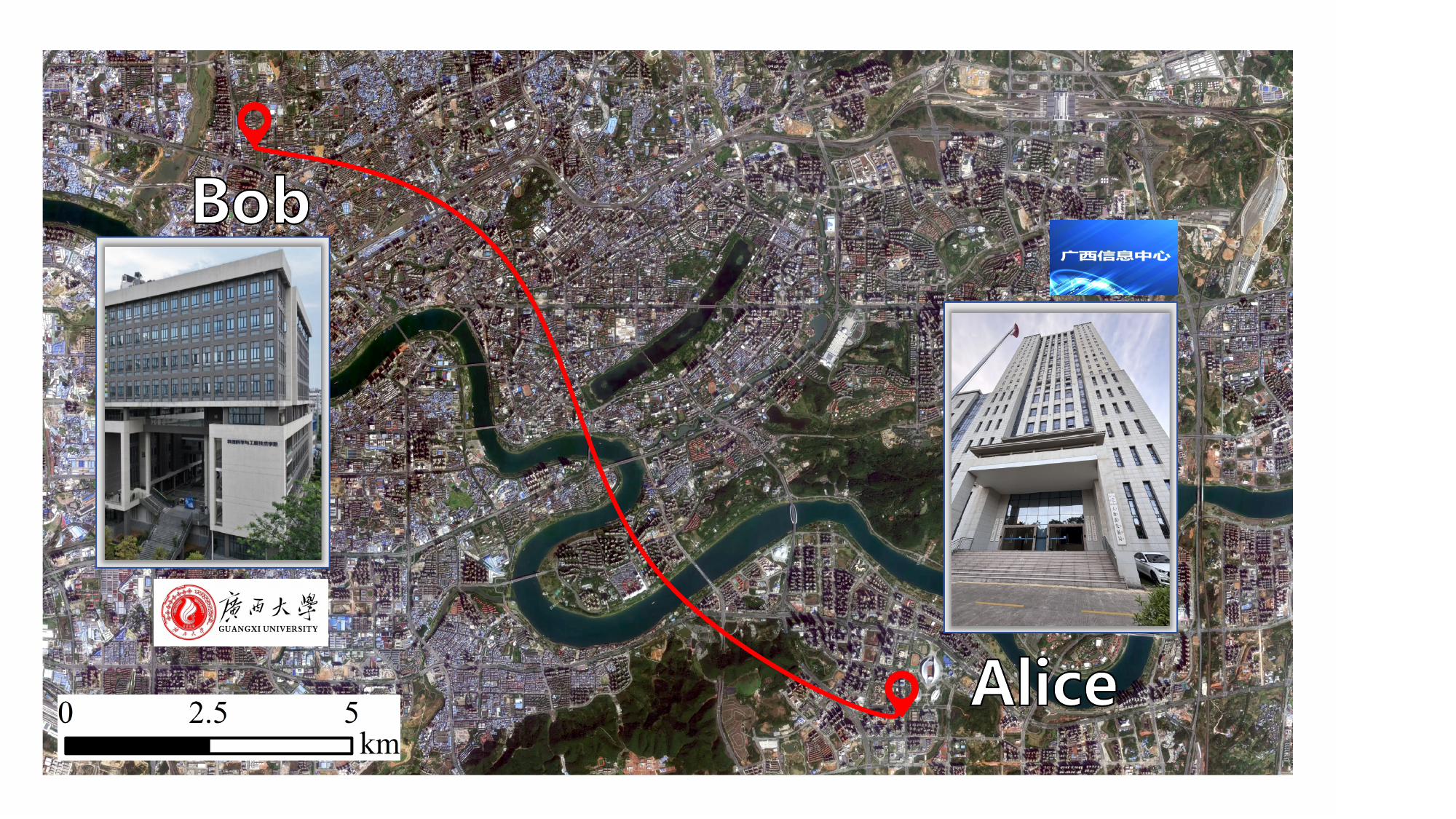}
		\caption{Aerial view of the fiber link used in the QKD field trial performed in Nanning, China. [©2025 Baidu]. The link connects the transmitter (Alice), located at the Guangxi Zhuang Autonomous Region Information Center, with the receiver (Bob), situated at the School of Physical Science and Technology, Guangxi University.}
		\label{fig_deployment_1}
	\end{figure*}
	
	
    The organizational structure of this paper is as follows. In Sec.~\ref{Experimental setup}, we describe the field-deployed QKD system and the corresponding experimental setup. In Sec.~\ref{Experimental Results}, we present the corresponding experimental results. Finally, we conclude the paper in Sec.~\ref{Conclusion}.

	\section{Setup}\label{Experimental setup}

	\begin{figure*}
		\centering
		\includegraphics[width=0.7\linewidth]{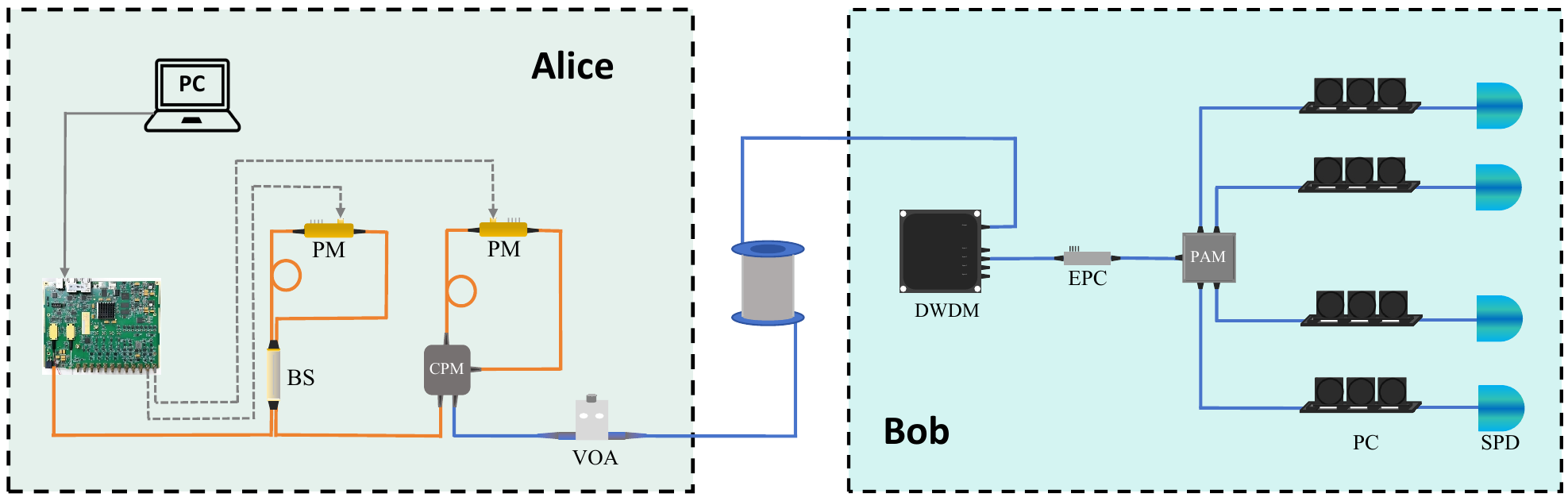}
		\caption{Schematic representation of the setup employed in the experiment. In the upper-left and upper-right boxes we represent the setup for the transmitter(Alice) and receiver(Bob), respectively. BS, beam splitter; PM, phase modulator; CPM, customized polarization module; VOA, variable optical attenuator; DWDM, dense wavelength division multiplexing; EPC, electronic polarization controller; PAM, polarization analysis module; PC, polarization controller; SPD, single photon detector; PM fiber, polarization-maintaining fiber; SM fiber, single mode fiber.}
		\label{fig_setup_2}
	\end{figure*}
	
	Figure \ref{fig_deployment_1} illustrates aerial view of  the experimental configuration, where the transmitter (Alice) is located at the Guangxi Zhuang Autonomous Region Information Center and the receiver (Bob) at the School of Physical Science and Technology, Guangxi University.  The total measured link loss is approximately 18 dB. The fiber link withstands environmental disturbances as it traverses the Yongjiang River and the bustling urban area. 
	
	The used experimental setup is a polarization-encoding QKD setup, as shown in Fig. \ref{fig_setup_2}. On the transmitter side (left panel of Fig. \ref{fig_setup_2}), a commercial laser generates an optical pulse train with a repetition rate of 100 MHz and a wavelength of 1550 nm. This pulse train is first encoded by an intensity modulator to prepare two intensities of signals required for the one-decoy-state BB84 protocol. The intensity modulator employs a Sagnac interferometer configuration\cite{2021MaOptLett,2019AgnesiOptLett,2018RobertsOptLett}, consisting of a 50:50 polarization-maintaining beam splitter (PMBS), a lithium niobate phase modulator (PM), and a 1 m polarization-maintaining fiber (PMF) delay line. Following intensity modulation, the optical pulses are immediately coupled into a Sagnac-based polarization modulator (Sagnac-POL) module for qubit polarization encoding. The core component of this module is a custom polarization beam splitter (CPM) based on a Sagnac interferometer configuration and a PM. A crucial design aspect is the precise alignment of the CPM's polarization-maintaining fiber input axis at a 45° angle relative to the principal axes of the beam splitter. This alignment enables efficient splitting of incident pulses into two orthogonally polarized components. By applying a pre-calibrated voltage signal from the RF output port on the 100 MHz board to the PM within the Sagnac-POL, we prepare the four BB84 polarization states as follows:
	
	\begin{equation}
		\ket{\psi} = \frac{\ket{H} + e^{i\theta}\ket{V}}{\sqrt{2}}, \quad \theta \in \left\{0, \frac{\pi}{2}, \pi, \frac{3\pi}{2}\right\}
		\label{eq:psi}
	\end{equation}
	where $\theta$ denotes the applied phase and $\theta \in \left\{0, \pi\right\}$ ($\theta \in \left\{\frac{\pi}{2}, \frac{3\pi}{2}\right\}$) corresponds to the states in the $Z$ ($X$) basis.
	
	Subsequently, the polarization-modulated light exits from the CPM's output port, is attenuated to the single-photon regime by variable optical attenuator (VOA), and is transmitted over the quantum channel.
	
	At the receiver (Bob), a dense wavelength division multiplexing (DWDM) filter suppresses dark count increases caused by light leakage and imperfect isolation in the field-deployed fiber link, reducing the system noise count from approximately 250 Hz to about 40 Hz, approaching the dark count level of the SPD. An electrically controlled polarization controller (EPC) then aligns the incoming polarization states with the measurement basis. Photon detection is achieved using four commercial superconducting nanowire single-photon detectors (SNSPDs, Photon Technology Co., Ltd.). The detectors operate at $2.1\,\text{K}$, with a detection efficiency of approximately $60\%$, a dead time of $40\,\text{ns}$, a timing jitter of $\leq 70\,\text{ps}$, and a dark count rate of about $40\,\text{Hz}$. A polarization controller (PC) is placed before each SNSPD to adjust the polarization of the incoming photons,  ensuring maximum detection efficiency. Detection events are recorded using a high-speed time-to-digital converter (TDC, TimeTagger20, Swabian Instruments) and transmitted to Bob's computer for time synchronization and polarization drift compensation.
	
	The time recovery employs the distributed frame synchronization technique proposed in~\cite{2025ChenJLT}, which reconstructs the clock cycle from event detection without relying on any external reference signal. A brief overview of this method is provided below. Specifically, Alice generates a correlation code $S^{A}$ of length $L$ and divides it into $L$ single-qubit segments. Each segment is concatenated with $M$ random qubits to form a composite qubit sequence, which is then transmitted to Bob through a lossy quantum channel. Upon receiving the qubit sequence, Bob performs period recovery, time compensation, and time offset estimation based on the detected qubit data.
	
	To compensate for polarization drift induced by environmental disturbance in deployed fiber channels, we employ a qubit-based polarization compensation scheme, similar to to the work in~\cite{Ding2017,Agnesi2020}. This method uses the shared qubit sequence between the communicating parties to realize dynamic polarization control. The procedure is as follows:

	After time synchronization, the QBER is calculated from the disclosed quantum bit sequence. The QBER is then fed into a gradient descent algorithm running on Bob’s computer to control the EPC for polarization compensation. The EPC consists of three fiber squeezers oriented $45^\circ$ from each other, which control the polarization of the transmitted light by applying stress and strain to the fiber. The feedback algorithm initially checks whether the QBER exceeds a threshold. If so, the three squeezers are adjusted in sequence. The algorithm first applies an initial adjustment to one of the squeezers along a certain direction. If the QBER decreases after the adjustment, it continues adjusting in that direction; otherwise, it immediately applies one reverse adjustment, ends the current round of adjustment for that squeezer, and then switches to the next squeezer. The above process is performed cyclically until the QBER falls below a preset threshold.

	\section{Experimental result}\label{Experimental Results}

    \begin{figure}[htbp]
    	\centering
    	\captionsetup[subfigure]{labelformat=empty}
    	
    	\begin{subfigure}[t]{0.5\linewidth}
    		\centering
    		\begin{tikzpicture}
    			\node[inner sep=0pt] (image) at (0,0) {\includegraphics[width=\linewidth]{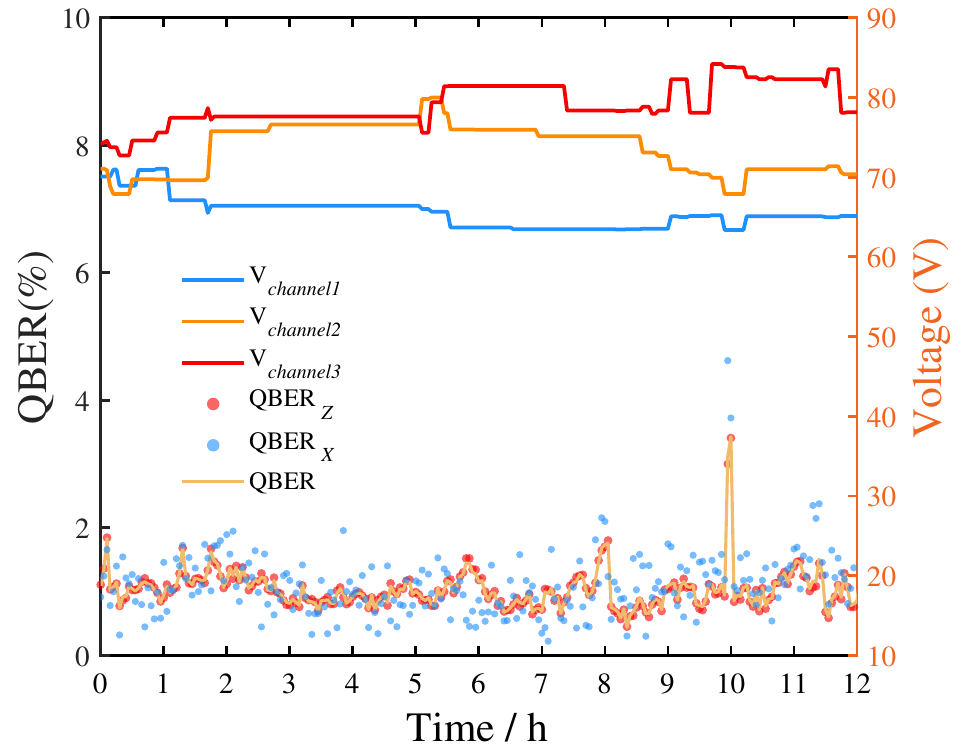}};
    			\node[anchor=north west, xshift=-6pt, yshift=-1pt] at (image.north west) {\textbf{(a)}};
    		\end{tikzpicture}
    		\caption{}   
    		\label{fig_volt_QBER12h_3}
    	\end{subfigure}
    	\hfill
    	\begin{subfigure}[t]{0.47\linewidth}
    		\centering
    		\begin{tikzpicture}
    			\node[inner sep=0pt] (image) at (0,0) {\includegraphics[width=\linewidth]{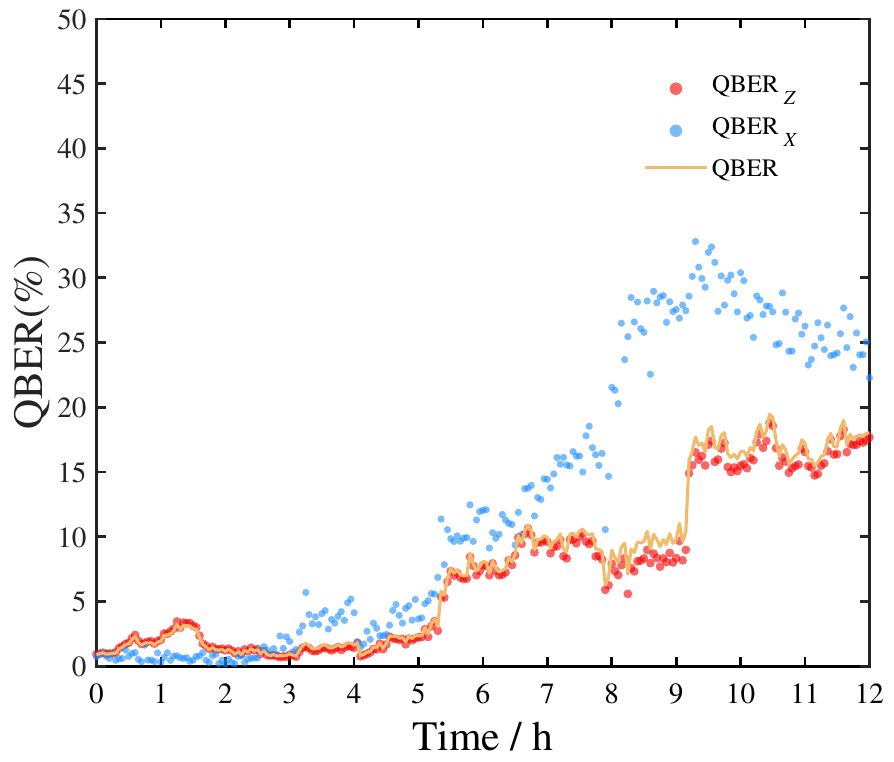}};
    			\node[anchor=north west, xshift=-6pt, yshift=-1pt] at (image.north west) {\textbf{(b)}};
    		\end{tikzpicture}
    		\caption{}
    		\label{fig_volt_QBER12h_withoutEPC}
    	\end{subfigure}
    	
    	\captionsetup[subfigure]{labelformat=default}
    	\vspace{-15pt}
    	\caption{The QBER of the system during a 12-hour continuous operation in a field-deployed fiber link with 18 dB channel loss is presented as follows: (a) The QBER for the $Z$ and $X$ bases (left axis) and the feedback voltage of the EPC (right axis, with channels 1, 2, and 3 in blue, purple, and orange, respectively), with polarization control. (b) The QBER for the $Z$ and $X$ bases without polarization control. The red (blue) points denote the quantum bit error in the $Z$ ($X$) basis, recorded every 2.5 minutes.}
    	\label{fig_combined_qber}
    \end{figure}
	
	We first evaluate the long-term stability of the QBER and the performance of real-time polarization feedback control. As shown in Fig.~\ref{fig_combined_qber}\subref{fig_volt_QBER12h_3}, we recorded the real-time QBER and corresponding feedback voltage of the EPC during a 12-hour continuous operation over a field-deployed fiber link with a channel loss of 18~dB. The feedback voltage of the EPC was adjusted in real time, and the QBER exhibited minor fluctuations due to environmental disturbances such as temperature variations and mechanical stress in the metropolitan fiber network. Despite these disturbances, the system maintained an exceptionally low average QBER. An average $\text{QBER}_Z$ is measured to be $1.12 \pm 0.51\%$ for the $Z$ basis, while for $X$ basis, the average $\text{QBER}_X$ is $1.10 \pm 0.51\%$. The total QBER for both two bases is $1.12 \pm 0.48\%$. Notably, around the 10th hour of operation, an abnormal increase in the QBER occurred, which is attributed to unpredictable factors such as external vibrations caused by subway operations and construction activities, which inevitably induce polarization disturbances in the optical fiber. It can be observed that the polarization feedback algorithm effectively compensates for these polarization drifts by adjusting the driving voltage of the EPC.
	
	For comparison, we turned off the polarization control and performed the same 12-hour continuous test, with an average QBER of $7.64 \pm 6.23\%$. As shown in Fig.~\ref{fig_combined_qber}\subref{fig_volt_QBER12h_withoutEPC}, the QBER begins to fluctuate one hour after system operation. From the 5th hour onward, the fluctuation amplitude increases markedly, and the average QBER thereafter reaches as high as nearly 20\%. This indicates that the introduction of active polarization feedback is essential, demonstrating the robustness of the proposed qubit-based synchronization and polarization compensation method.
		
	We further evaluate the secure key rate under various transmission loss conditions. The experiment was first conducted over a deployed fiber link with an optical loss of approximately 18~dB. To assess performance over longer distances, VOA was used to emulate higher channel losses of 30~dB and 40~dB. For link losses of 18~dB and 30~dB, the total length of the periodic correlation code was set to $L$, and the ratio $M$ was configured as 7:1. Under the 40~dB loss condition, the ratio $M$ was adjusted to 3:1 to increase the number of synchronization sequences and reduce the risk of synchronization failure caused by excessive attenuation of the synchronization signal. For each distance, approximately $10^7$ detection events were accumulated. Before each estimation, a full optimization of the implementation parameters for the one-decoy-state protocol was performed to ensure optimal system performance~\cite{Li2022ddi}. For instance, in the 18~dB experimental scenario, the signal and decoy state intensities were set to $\mu = 0.565$ and $\nu = 0.143$, respectively, while the probabilities of sending the signal state and selecting the $Z$ basis were $p_\mathrm{\mu} = 0.798$ and $p_\mathrm{Z} = 0.944$. On Bob’s side, the probabilities of selecting the $Z$ and $X$ bases were fixed and equal.
	
	The finite-key analysis was performed using the method described in~\cite{rusca2018finite}, and the final secure key rate was estimated using the following formula:
	
	\begin{flalign}
		R = \biggl( s_{z,0}^L + s_{z,1}^L & \left( 1 - h(e_{z,1}^{ph}) \right) - \lambda_{EC} - 6 \log_2 \frac{19}{\epsilon_{sec}} - \log_2 \frac{2}{\epsilon_{cor}} \biggr) \cdot \frac{q \cdot f}{N} 
	\end{flalign}
	
	Here, $s_{z,0}^L$ denotes the lower bound of detection events that Bob receives when Alice sends a vacuum state in the $Z$ basis. $s_{z,1}^L$ denotes the lower bound of single-photon detection events in the $Z$ basis, assuming that Alice sends only single-photon states. $e_{z,1}^{ph}$ represents the upper bound of the phase error rate for single-photon events. $\lambda_{EC} = n_z f_e h(e_z)$ denotes the number of bits consumed during error correction, where $e_z$ is the quantum bit error rate (QBER) and $f_e = 1.16$ is the error correction efficiency factor. $\epsilon_{sec}$ and $\epsilon_{cor}$ are security and correctness parameters, respectively. The binary entropy function is defined as $h(x) = -x \log_2(x) - (1-x) \log_2(1-x)$. $f$ is the system repetition frequency, and $N$ denotes the total number of optical pulses, including both synchronization and random pulses. $q = $M$ / ($M$+1)$ denotes the fraction of signals used for key distribution.
	
\begin{figure}
		\centering
		\includegraphics[width=0.72\linewidth]{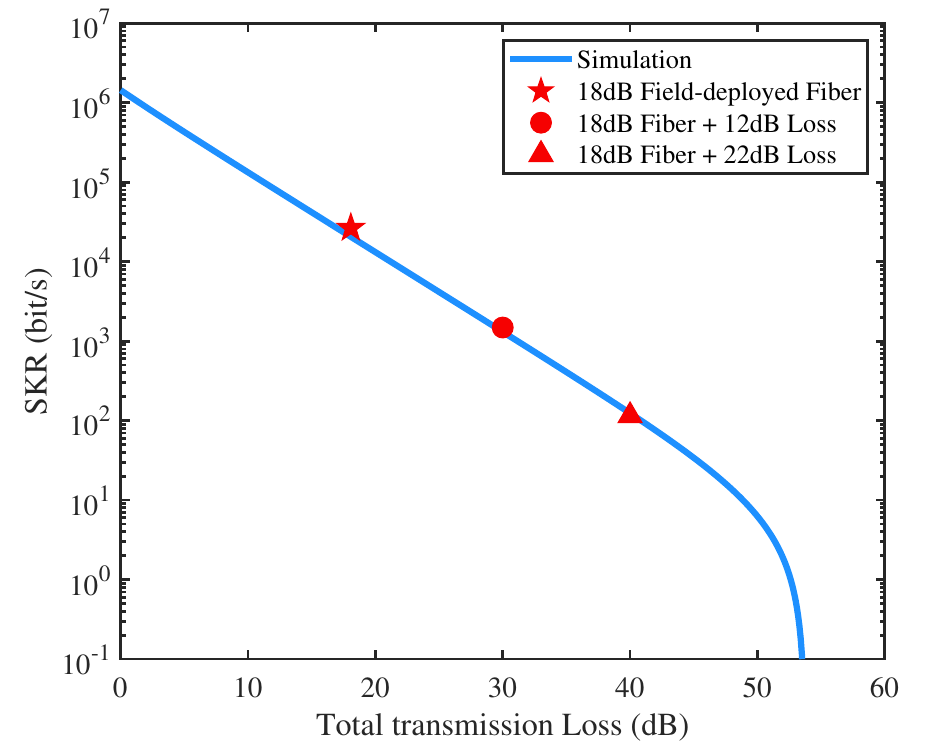}
		\caption{The secure key rate under different transmission losses. The blue line shows the simulation results based on our experimental parameters, the red stars represent the experimental results obtained over real fiber links. The red circle and triangle markers denote the results measured over the same field fiber with additional 12 dB and 22 dB losses emulated by VOA, corresponding to total losses of 30 dB and 40 dB, respectively.}
		\label{fig_SKR_3}
	\end{figure}

	The experimental results are presented in Fig.~\ref{fig_SKR_3}, and the corresponding parameters and performance metrics are summarized in Table~\ref{table:detailed_results}. The results clearly demonstrate that the system sustains a high secure key rate (SKR) of 26.6 kbit/s over the deployed fiber link in Nanning. Even under extreme channel attenuations of 30 dB and 40 dB—emulated using VOA—the system achieves finite-key secure rates of 1.48 kbit/s and 115 bit/s, respectively. To demonstrate our experimental results, we compare our work with several recent field experiment results in Table~\ref{tab:field_comparison}. The comparison shows that under similar channel loss conditions, our system achieves a lower QBER. Notably, compared with other works that also employ qubit-based synchronization, our work obtains a higher SKR (26.6 kbit/s vs. 11.5 kbps) under a higher channel loss (18 dB vs. 9 dB). These findings exhibit strong agreement with theoretical predictions, confirming the robustness and efficiency of the qubit-based synchronization and polarization control mechanisms under high-loss conditions.
	
\begin{table*}[!ht]
    \begin{minipage}{\textwidth} 
	\centering
	\captionsetup{width=\textwidth, justification=justified, singlelinecheck=false}
	\caption{Experimental parameters and results. Loss is the channel losses, $\mu$ ($\nu$) is the intensity of the signal (decoy) state, $P_{\mu}$ ($P_{\nu}$) is the selection probability of the signal (decoy) state, $P_{z}$ ($P_{x}$) is the probability of choosing $Z$ ($X$) basis, $n_{z}$ is the amount of data accumulated on the $Z$ basis, $t$ is the total accumulated time of measurement, Ratio denotes the ratio between the number of qubits for synchronization and that of random polarization states of four BB84 polarizations. $\tau_b$ is the period measured by Bob, $n_{z,\mu}$ ($n_{x,\mu}$) is the raw count measured under $Z$ ($X$) basis for the signal state, $n_{z,\nu}$ ($n_{x,\nu}$) is the raw count measured for the decoy state under $Z$ ($X$) basis, $m_{z,\mu}$ ($m_{x,\mu}$) is the error count measured under $Z$ ($X$) basis for the signal state, $m_{z,\nu}$ ($m_{x,\nu}$) is the error count measured for the decoy state under $Z$ ($X$) basis, QBER$_{z}$ is quantum bit error rates of $Z$ basis, $\phi_z^{U}$ is the upper bound of phase error rate, $s_{z,1}^{L}$ is the lower bound for measuring single-photon events in the $Z$-basis, $l$ is the secret key length, and SKR is the final secure key rate.}
	\label{table:detailed_results}
	
	\setlength{\tabcolsep}{3pt}
	
	\begin{tabular*}{\textwidth}{@{\extracolsep{\fill}}cccccccccc}
		\toprule
		Loss (dB) & $\mu$ & $\nu$ & $P_{\mu}$ & $P_{\nu}$ & $P_z$  & $\tau_b$ (ns) & $n_z$ & Ratio\\
		\midrule
		18 & 0.466 & 0.127 & 0.761 & 0.239 & 0.935 & $10.0 \pm 8.8 \times 10^{-8}$ & 9,921,359 & 1:7\\
		30 & 0.465 & 0.126 & 0.761 & 0.239 & 0.934 & $10.0 \pm 3.2 \times 10^{-7}$ & 9,844,391 & 1:7\\
		40 & 0.464 & 0.125 & 0.761 & 0.239 & 0.931 & $10.0 \pm 1.8 \times 10^{-7}$ & 9,898,948 & 1:3\\
		\bottomrule
	\end{tabular*}
	
	\vspace{0em}
	\vspace{0em}
	
	\begin{tabular*}{\textwidth}{@{\extracolsep{\fill}}cccccccccc}
		\toprule
		Loss (dB) & $n_{z,\mu}$ & $m_{z,\mu}$ & {$n_{x,\mu}$} & {$m_{x,\mu}$} & $n_{z,\nu}$ & $m_{z,\nu}$ & $n_{x,\nu}$ & $m_{x,\nu}$ & $\phi_z^{U}$ \\
		\midrule
		18 & 9,157,392 & 84,469 & 735,824 & 6,302 & 763,967 & 7,293  & 61,775 & 762  & 0.0227\\
		30 & 9,086,317 & 102,679 & 730,681 & 9,157 & 758,074 & 8,919  & 61,348 & 1,151 & 0.0323\\
		40 & 9,115,411 & 150,978 & 766,631 & 15,029 & 783,537 & 22,114 & 68,149 & 1,969 & 0.0443\\
		\bottomrule
	\end{tabular*}
	
	\vspace{0em}
	\vspace{0em}
	
	\begin{tabular*}{\textwidth}{@{\extracolsep{\fill}}ccccccc}
		\toprule
		Loss (dB) & $s_{z,1}^{L}$ & $l$ & $t$ (s) & $\tau_b$ (ns) & QBER$_z$ (\%) & SKR (bit/s) \\
		\midrule
		18 & 5,620,092 & 3,868,037 & 145.5    & $10.0 \pm 8.8 \times 10^{-8}$ & 0.925 & $2.66 \times 10^{4}$ \\
		30 & 5,533,409 & 3,370,711 & 2,271.9  & $10.0 \pm 3.2 \times 10^{-7}$ & 1.134 & $1.48 \times 10^{3}$ \\
		40 & 5,528,936 & 2,622,953 & 22,779.7 & $10.0 \pm 1.8 \times 10^{-7}$ & 1.749 & $1.15 \times 10^{2}$ \\
		\bottomrule
	\end{tabular*}
    \end{minipage}  
\end{table*}
	
	\begin{table}[H]
		\centering
		\captionsetup{width=\textwidth, justification=justified, singlelinecheck=false}
		\footnotesize   
		\setlength{\tabcolsep}{3.3 pt}   
		
		\caption{Comparison of recent QKD field tests using the BB84 protocol. Sync is the synchronization method used in each system. QBER$_{Z}$ and QBER$_{X}$ is quantum bit error rates of $Z$ ($X$) basis. Since some parameters are not explicitly provided in the references, they are indicated by a dash “-”.}
		\label{tab:field_comparison}
		\begin{tabular}{l c c c c c c c c c}
			\toprule
			References & Sync & Clock rate & Loss(dB) & Distance(km) & QBER$_Z$ & QBER$_X$ & QBER & SKR\\
			\midrule
			Bacco et al.~\cite{2019Bacco} & Classical Light & 595 MHz & 21 & 40 & 1.11\% & 3.93\% & - & 44.2 kbit/s\\
			Zahidy et al.~\cite{zahidy2024quantum} & Classical Light & 72.6 MHz & 9.6 & 18.1 & - & - & 3.25\% & 47.9 kbit/s\\
			Wu et al.~\cite{wu2025integration} & Classical Light & 595 MHz & - & 25.2 & - & - & 2.10\% & 0.41 kbit/s \\
			Avesani et al.~\cite{2021Avesani} & Qubit-based & 50 MHz & 9 & 3.4 & 2.00\% & 1.10\% & - & 11.5 kbps\\
			This work & Qubit-based & 100 MHz & 18 & 30.7 & 1.10\% & 1.10\% & 1.12\% & 26.6 kbit/s\\
			\bottomrule
		\end{tabular}
	\end{table}

	\section{Conclusion and discussion}\label{Conclusion}
    This study successfully demonstrates the long-term stability and robustness of a qubit-based distributed frame synchronization QKD system deployed over a real-world metropolitan fiber network. By integrating polarization feedback control and qubit-based synchronization, the system achieves stable synchronization and real-time polarization compensation without the need for external classical channels or additional hardware. Experimental results confirm its excellent performance under challenging conditions: during 12 hours of continuous operation, the system maintained an average QBER of $1.12 \pm 0.48\%$ at 18 dB channel loss, while achieving a secure key rate of 26.6 kbit/s. Even under extreme attenuation of 40 dB, the system sustained a finite-key secure rate of 115 bit/s, demonstrating remarkable tolerance to high channel loss. These findings underscore the practicality and field readiness of qubit-based synchronization for real-world QKD deployments, where environmental fluctuations are unavoidable.

    Future research could overcome the practical constraints inherent to the used qubit-based distributed frame synchronization methods in real-world applications. For example, the algorithm adopted in this work requires high-speed sampling of SPD counts and the execution of FFT operations during the frequency recovery process (to satisfy the Nyquist sampling theorem). This imposes high demands on data throughput and computational speed in the post-processing stage, which may introduce system latency. To address this, more efficient clock recovery algorithms~\cite{lu2025fast} could be considered to reduce latency. Furthermore, because the synchronization algorithm requires extracting a portion of the quantum states as a synchronization sequence to perform synchronization, it inevitably reduces the SKR. Therefore, time synchronization techniques that further integrate Bayesian methods~\cite{cochran2021} could be considered to achieve remote clock synchronization without sacrificing synchronization qubits. These improvements are expected to further enhance overall system performance and extend the approach to more practical application scenarios.

	\FloatBarrier
	\begin{backmatter}
		
		\bmsection{Funding} National Natural Science Foundation of China (Nos. 62171144, 62031024, and  11865004), Guangxi Science Foundation (Nos.2021GXNSFAA220011, 2021AC19384 and 2025GXNSFAA069137), Open Fund of IPOC (BUPT) (No. IPOC2021A02), Guangdong Basic and Applied Basic Research Foundation (2024B1515120030), and Bagui Scholars Programme (W.X.-G., GXR-6BG2424001). 
		
		\bmsection{Disclosures} The authors declare no conflicts of interest.
		
		\bmsection{Data Availability}Data underlying the results presented in this paper are not publicly available at this time but may be obtained from the authors upon reasonable request.
	\end{backmatter}

	\bibliography{Ref}
		
\end{document}